# Possible Magnetic Behavior in Oxygen-deficient β-PtO$_2$

Yong Yang[1], Osamu Sugino[1,2], and Takahisa Ohno[1]

1. *National Institute for Materials Science, 1-2-1 Sengen, Tsukuba, 305-0047, Japan*

2. *Institute for Solid State Physics, University of Tokyo, 5-1-5 Kashiwanoha, Kashiwa 277-8581, Japan*

We studied the electronic properties of beta-platinum dioxide (β-PtO$_2$), a catalytic material, based on density functional theory. Using the GGA+U method which reproduces the GW band structures and the experimental structural parameters, we found that the creation of an oxygen vacancy will induce local magnetic moment on the neighboring Pt and O atoms. The magnetism originates not only from the unpaired electrons that occupy the vacancy induced gap state, but also from the itinerant valence electrons. Because of antiferromagnetic (AF) coupling and the localized nature of gap states, the total magnetic moment is zero for charge-neutral state ($V_O^0$) and is ~ 1 μ$_B$ for singly-charged states ($V_O^\pm$). Calculation of grand potential shows that, the three charge states ($V_O^0$, $V_O^\pm$) are of the same stability within a small region, and the negatively charged state ($V_O^-$) is energetically favored within a wide range of the band gap. On this basis we discussed the implication on catalytic behavior.

I.  INTRODUCTION

Vacancy-induced magnetism has been reported in recent years in semiconducting and insulating materials, ranging from monoxide (CaO [1], MgO [2, 3], ZnO [4-6]), dioxides ($HfO_2$ [7, 8], $SnO_2$ [9], $TiO_2$ [10-14]) to wide-gap III nitrides (GaN and BN [15]). Similar magnetic phenomenon is also found in carbon systems such as diamond, graphite and graphene [16-20]. The common feature is that the defect-free materials are nonmagnetic whereas local magnetic moment is induced around the vacancy sites and interacts with each other to cause the material to become magnetic. Such magnetism is *intrinsic* because no foreign species are introduced into the host, and has thus attracted attention as a new way to synthesize dilute magnetic semiconductors. In oxides and nitrides, the physical origin is generally attributed to partial occupation of the dangling $2p$ orbitals or the open $d$ shells around the cation/anion vacancies. Although the mechanism regarding the vacancy-induced magnetism in carbon materials seems somewhat complex [16-20], the role of dangling bonds is evident in the appearance of local magnetic moment [19]. In this work, we show that the vacancy-induced intrinsic magnetism will possibly present in the β phase of platinum dioxide (β-$PtO_2$).

Platinum oxides have been attracting technological attention as a catalyst for organic synthesis [21, 22] and as an electrode material for electronic devices [23]. More recently, attention was paid on surface oxides formed on platinum because of their role in catalyzing reduction and oxidation of adsorbates. The surface oxides are formed on Pt either at oxygen rich conditions under the ultra-high vacuum (UHV) [24-29] or at high potentials imposed electrode-solution interface [30, 31]. Under electrochemical conditions, the Pt electrode is oxidized when the electrode potential is above 0.7 V vs. the reversible hydrogen electrode (RHE) [32]. Imai *et al.* [33] found that thin film comprised of β-$PtO_2$ is formed when the potential is 1.4 V vs. RHE. These experiments indicate the fuel cell reactions such as the oxygen reduction reaction (ORR), cannot be well understood unless elucidating the property of the oxide: In particular, the electronic property of $PtO_2$ is rather different from Pt in that the dioxide is insulating. One may wonder how the charged carriers can reach the reaction center without being blocked by the dioxide film. The experiment of Sun *et al.* [34] was motivated by this question, in which they investigated the defects distribution, such as oxygen vacancies and interstitials. Yet, the role of the defects in carrier transfer remains unclear. In this context, characterizing the electronic energy level of the defects in β-$PtO_2$ is an important step forward, and such study is lacking so far.

Although it has been more than 80 years since its synthesis, the electronic structure of $PtO_2$ as a transition metal oxide is not well understood. For instance, experimentally measured band gap for α-$PtO_2$ ranges from 1.3 eV [35] to 1.8 eV [36] while no experimental band gap is available for β-$PtO_2$. Using density functional theory (DFT) calculations within the local density approximation (LDA), Wu and Weber investigated the stability of the β-$PtO_2$ crystal structure [37]. However, their calculated electronic density of states (DOS) indicates a metallic character of β-$PtO_2$, which contradicts with the experimental measurement that β-$PtO_2$ is a semiconductor with a resistivity of ~ $10^6$ Ω·cm [38]. Experiments show that amorphous $PtO_2$ film is a semiconductor whose resistivity is dependent on deposition methods [39]. With the presence of oxygen defects, the physical and chemical properties of an oxide can be significantly changed and may exhibit novel properties such as magnetism that is absent in the defect-free system. It is unknown what new properties can be expected when oxygen vacancies are introduced into $PtO_2$.

Here, we investigate the electronic structures of isolated oxygen vacancies in $\beta$-$PtO_2$ using first-principles calculations based on density functional theory (DFT). The strong correlation effect of valence electrons is taken into account by using the GGA+U scheme. The validity of GGA+U method is justified by GW calculations [40, 41], state-of-the-art the most accurate method for the calculating band gap in semiconductors and insulators. Creation of oxygen vacancy is found to induce local magnetic moment around the vacancy site, which is not only due to the unpaired excess electrons that occupy the gap states but also the magnetic moment of itinerant valence electrons. The magnetism is characterized by analyzing the electron spin distribution. We have also studied the stability and magnetization of singly-charged states and discuss the possible implication on catalysis.

In the following sections, we will first briefly describe the computational and modeling methods employed in this study. The results and discussion regarding the electronic and magnetic properties of oxygen-deficient $\beta$-$PtO_2$ at different charge states are presented in Sec. III. The concluding part is given in Sec. IV.

## II. Computational & Modeling Methods

$\beta$-$PtO_2$ has a $CaCl_2$-type crystal structure [42], which is similar to rutile $TiO_2$ but with a slight distortion in the direction of cell axes *a* and *b* (Fig. 1(a), Table I). Besides, it is the only known group VIII metal dioxide that has a CaCl$_2$-type crystal structure [43]. Shown in Fig. 1(a), is the primitive cell of β-PtO$_2$, containing two Pt atoms and four O atoms. To study the oxygen vacancy state, we employed a (2×2×3) and (3×3×4) supercell of β-PtO$_2$. Here we will focus on isolated vacancy (or a dilute state of them) in bulk β-PtO$_2$ although the clustering of vacancies is possible. The calculations were carried out by the Vienna *ab initio* simulation package (VASP) [44, 45], using a plane wave basis set and the PAW potentials [46, 47]. The exchange-correlation interactions were described by the generalized gradient approximation (GGA) with PBE functional [48]. The strong on-site Coulomb interaction was treated using the simplified GGA+U method introduced by Dudarev *et al.* [49]. As for the GW calculations, we take the conventional GW approach [41] implemented in VASP code [44, 45], the so-called G$_0$W$_0$ method to calculate the energy spectrum of quasiparticle. The energy cutoff for plane waves is 600 eV. For the GGA and GGA+U calculation of primitive unit cell, we used a 16×16×16 k-mesh. For the (2×2×3) supercell (with and without oxygen vacancy), a 3×3×3 k-mesh was employed for structural relaxation and a 4×4×4 k-mesh was used for the calculation of electronic density of states (DOS) and charge density. For the (3×3×4) supercell, a 2×2×2 k-mesh was employed for all the calculation. In the GW calculations, because of the heavy computational burden, the primitive unit cell is considered only and the employed k-meshes range from 2×2×2 to 6×6×6, to test the convergence of obtained band gap. The k-meshes were generated by the Monkhorst-Pack scheme [50]. The tetrahedron method with Blöchl corrections [51] was used for integral in Brillouin zone (BZ).

Starting with the experimentally determined structure of β-PtO$_2$ [42], we optimized the atomic position and cell geometry using DFT calculations. Because many transition metal oxides belong to strongly-correlated electron system [52], both GGA and GGA+U method were employed to make a comparison. Since no experimental data were available for band gap, the data we can rely on are the lattice parameters. In GGA+U calculation, the effective on-site Coulomb repulsion (U$_{eff}$) adopted for the 5$d$ orbital of Pt is 7.5 eV, which gives the best lattice parameters comparing to experiments. The obtained cell parameters and the corresponding experimental data are listed in Table I. For GGA and GGA+U calculation, the unit cell volume is enlarged by ~ 5% and ~ 2%, respectively. The relative size of cell axes *a* and *b* given by GGA+U calculation is in agreement with experiment (*a* < *b*) while it is in reverse order by GGA method (*a* > *b*). The length of *c* axis given by GGA+U is also in better agreement with experiment. Figure 1(b) shows the calculated electronic DOS for the valence and conduction band of β-PtO$_2$ with GGA and GGA+U method, respectively. The band gap given by GGA+U is in good agreement with the experimental value of amorphous PtO$_2$ (~ 1.20 eV) [53], while the band gap predicted by GGA is much smaller (~ 0.46 eV). This is due to the fact that LDA/GGA calculations usually underestimate the band gap of semiconductors and insulators. Moreover, subsequent GW calculation predicts that the band gap of β-PtO$_2$ is ~ 1.2 eV, which further demonstrates the validity of the U$_{eff}$ value employed here. Considering the fact that β-PtO$_2$ has a larger resistivity than α-PtO$_2$ (10$^6$ Ω·cm vs 10$^4$ Ω·cm) [38, 54, 55], it would be reasonable that their band gap is of similar magnitude. Therefore, the GGA+U method with U$_{eff}$ = 7.5 eV (Pt 5$d$) was employed in all the following calculations for systems with oxygen vacancies. Figure 1(c) plots the GGA+U total energy as function of unit cell volume. The data were least-squared fitted to the Murnaghan equation of state [56, 57]. The bulk modulus ($B_0$) and its pressure derivative ($B_0'$) at equilibrium volume were deduced to be 208.2 GPa and 5.1, respectively. The calculated GGA+U band structure of bulk $\beta$-PtO$_2$ is shown in Fig. 1(d). It is clear that the band gap is indirect.

Figure 2(a) shows the calculated electronic density of states (DOS) of $\beta$-PtO$_2$ using GGA, GGA+U, and the conventional GW ($G_0W_0$) method. The calculated band gap by GGA, GGA+U and $G_0W_0$ is 0.46 eV, 1.20 eV, 1.30 eV, respectively. The GGA band gap is very close to the value given by a previous DFT calculation [58], which is 0.43 eV. A 4×4×4 k-mesh and an energy cut-off of 600 eV for plane waves and totally 400 energy bands are used for the $G_0W_0$ calculation. The GGA energy bands along some lines joining the symmetry points in the k-space are shown in Fig. 2(b), together with the $G_0W_0$ quasiparticle energies. One sees that the energy dispersion of GGA and $G_0W_0$ data points is similar in the k-space. The band gap of this GGA energy band lines looks like ~ 0.7 eV, a little bit larger than the value obtained from DOS calculation. This is due to the usage of different k-mesh, the electronic states corresponding to the band gap of ~ 0.46 eV are missing from data points on the high symmetry lines. Compared to the GGA bands, the major difference is that the $G_0W_0$ energies at the conduction bands are shifted up to higher values. Calculations using denser k-meshes, higher energy cutoff of plane wave and larger number of energy bands show that, the band gap converges to ~ 1.2 eV. More computational details can be found elsewhere [59].

The properties of an isolated oxygen vacancy were modeled by removing one oxygen atom in (2×2×3) and (3×3×4) supercells of β-PtO$_2$. The vacancy concentration is ~ 2.1% in a (2×2×3) supercell and ~ 0.69% in a (3×3×4) supercell. The corresponding stoichiometric formula for the two oxygen-deficient systems can be termed as PtO$_{1.958}$ and PtO$_{1.986}$, respectively. During structural optimization, both the atomic positions and the supercell geometries are fully relaxed. The oxygen vacancy formation energy ($E_{vf}$) was calculated as

$$E_{vf} = E\,[\text{PtO}_{2-\delta}] + 0.5E\,[\text{O}_2] - E\,[\text{PtO}_2] - T\Delta S, \qquad (1)$$

where the first three terms are the total energies of the oxygen-deficient β-PtO$_2$, gas phase O$_2$ (spin triplet state), and defect-free β-PtO$_2$, respectively. The value δ = 0.042 for the PtO$_{1.958}$ and δ = 0.014 for the PtO$_{1.986}$ system. The term Δ$S$ is the entropy change associated with creation of the vacancy. Since contribution from solid states is negligible compared to that from gaseous states, Δ$S$ is given by the entropy of the gaseous O molecules: We used the value 205.15 J·K$^{-1}$·mol$^{-1}$ measured at standard condition [60].

To analyze the redistribution of charge density upon vacancy creation, we computed electron density difference as

$$\Delta\rho = \rho[\text{PtO}_{2-\delta}] + \rho[\text{O}] - \rho[\text{PtO}_2], \qquad (2)$$

where $\rho[\text{PtO}_{2-\delta}]$ and $\rho[\text{PtO}_2]$ are the total electron density of the system with and without oxygen vacancy, respectively. The term $\rho[\text{O}]$ is the electron density of an isolated O atom centered at the vacancy site. Since an oxygen atom has two possible spin states in gas phase: triplet state and singlet state, we will adopt the $\rho[\text{O}]$ of both spin states for the calculation.

## III. Results and Discussion

### A. Charge-neutral system

We begin by studying the charge-neutral vacancy before turning to the charged ones. This is done by showing the formation energy, the density of states, the charge density difference and spatial distribution of the magnetization. The properties of a charge-neutral vacancy were studied by removing one oxygen atom from a (2×2×3) supercell of β-$PtO_2$, i.e., the $PtO_{1.958}$ system. Compared to defect-free system, the structural distortion is small: It is less than 0.3% for the lengths of lattice vectors, and less than 6% and 1% for the bond lengths and bond angles around the vacancy site, respectively.

Removal of an O atom from oxides generally leaves excess electrons around the vacancy site and modifies charge and spin distribution. Therefore, both spin-polarized and spin-unpolarized solutions were considered in our calculations. For the spin-unpolarized solution, the formation energy is calculated to be 2.45 eV. This energy shows no dependence on vacancy sites though there are four inquivalent O sites, mainly due to the fact that all the O atoms are coordinated by three Pt atoms with the same local bonding geometries. In spin-polarized solution, the total energy (and thus the vacancy formation energy) is further lowered by ~ 0.25 eV. Therefore the ground state of the oxygen-deficient system favors a spin-polarized solution. The magnetization density is spatially dependent and mainly localized at Pt atoms near the vacancy site, and the total magnetic moment is zero when charge-neutral. This indicates the local magnetic moments are interacting in an antiferromagnetic way. Major results are found the same when using a larger supercell (e.g., 2×2×4, 3×3×4) to model the isolated oxygen vacancy.

Left panel of Fig. 3 shows the spin-resolved total and partial density of states (DOS) of $PtO_{1.958}$. Near the top of valence band, the DOS is primarily contributed from the $2p$ orbitals of O atoms and secondarily from the $5d$ orbitals of Pt atoms, while opposite is true near the bottom of conduction band. Three vacancy states appear in the band gap: one spin-up and two spin-down. All the vacancy states show the hybridization of O $2p$ and Pt $5d$ orbitals. Two vacancy states, one spin-down (peak **a**) and one spin-up (peak **b**) locate just above the top of valence band. The other vacancy state (peak **c**) sits near the bottom of conduction band. For charge-neutral system, the vacancy state **a** is filled by one electron, and the other two vacancy states are empty. Each vacancy state can accommodate one electron at most. The band structures for spin-up and spin-down states are displayed in the right panels of Fig. 3. Judging from their small energy dispersion in k-space, and their sharp and isolated peaks in the DOS diagram, the vacancy states should be localized states. Indeed, this is demonstrated below by their electron density distribution. Results in a (3×3×4) supercell are similar, except that the relative intensity of the DOS peaks of vacancy states is smaller because of the lower vacancy concentration (~ 0.69%).

From Fig. 3, one can also find that the energy difference between the Fermi level and the lowest unoccupied vacancy state (peak **b** in left panel) is very small (~ 31 meV). That means a small perturbation may facilitate the electrons to jump to the vacancy state **b**. Indeed, we have found that in a slightly modified defect structure with slightly higher total energy (differs by ~ 9 meV), the Fermi level is found to go across the two vacancy states **a** & **b** and the system exhibits weak metallic characteristics, as shown in Fig. 4. This implies that the oxygen-deficient system is at the margin of switching between insulating and weak metallic state.

The isosurfaces of electron density of the vacancy states **a**, **b**, and **c** ($\rho_\mathbf{a}$, $\rho_\mathbf{b}$, $\rho_\mathbf{c}$, weighted average of all k-points) are displayed in Figs. 5(a)-(c). For all the panels, the major part of electron density (therefore electron wavefunction) is localized on the atoms around the vacancy site (referred to as Vo hereafter). Using the functionality of site- and *lm*-projection implemented in VASP code [61], we made analysis on the wavefunction characteristics of the vacancy states. The *lm*-component of each band is obtained by projecting it onto spherical harmonics within spheres of a radius $R_c$ around each atom and then the results were summed over all the k-points and atoms. The radius $R_c$ is 1.30 Å for Pt and 1.28 Å for O, which is approximately the covalent radius. The *lm*-component of each vacancy state is summarized in Table II. From Table II, one sees that vacancy state **a** largely consists of 2*p* components of O, with relatively smaller 5*d* components of Pt. For vacancy states **b** and **c**, the orbital hybridization between O and Pt atoms is strong and both the 2*p* and 5*d* components are significant. The sum of *lm*-component for vacancy states **a**, **b**, and **c** is 0.930, 0.931, and 0.831, respectively. All the numbers are close to but smaller than 1 (number of electron in each state), which means that the *lm*-projection has captured the major part but still lacks some of the characteristics. The missing features of wavefunction are due to the following two reasons: 1) The total volume occupied by the spheres centered at Pt and O atoms is ~ 81.2% of the supercell volume; 2) The basis set employed for projection is not complete. Variation of $R_c$ within a reasonable range will lead to small changes of the numbers in Table II, but the order of quantitative contribution from each orbital is kept.

Figure 6 shows the spatial distribution of magnetization density, which is the difference between spin-up and spin-down electron density. The three-dimensional (3D) isosurface is shown in Fig. 6(a), and the two-dimensional (2D) contours in the plane cutting through the three Pt atoms and the vacancy site, are shown in Fig. 6(b). The major part of magnetization is localized around Vo, especially around its three nearest neighboring Pt atoms (labeled as A, B, C). From Fig. 6(a), one sees that the magnetization density has a dumbbell shape aligned in parallel to the line connecting Pt and Vo. The difference is the direction of the magnetization: Both A and B are spin-up while C is spin-down. The overall feature is that the local magnetic moments of the three Pt atoms form an isosceles triangle arrangement around the Vo site.

By integrating the electron density inside a sphere of radius 1.30 Å centered at each Pt, the number of spin-up (spin-down) electrons at A, B, C is counted to be 4.57 (3.98), 4.57 (3.99), 3.91 (4.58), respectively. The local magnetic moment is calculated as a difference between the spin-up and spin-down electrons, yielding 0.59 $\mu_B$ (Bohr magneton), 0.58 $\mu_B$, and −0.67 $\mu_B$ for A, B, and C, respectively. Within computational error bar, the magnetic moment of Pt A and B is the same. This reflects the geometric equivalence of A and B and inequivalence of C from A and B. Indeed, the line connecting Pt and Vo is oriented along inequivalent direction: it is along [11-1] and [111] directions (which are equivalent in real space) for A and B, respectively, but it is along [110] direction for C. The angle Vo-Pt-O formed by Vo, the Pt atom (A, B, C) and the nearest O atom of Vo are also inequivalent: The angle is 77.91° for A and B, while it is 87.11° for C.

To further analyze the characteristics of magnetization, we computed the *lm*-component of each local magnetic moment using similar method as the analysis of vacancy states. The *lm*-component of each band was obtained by projection onto spherical harmonics around each atom and the results were summed over all the filled bands. The major part of local magnetic moment of Pt atom A is contributed from the 5*d* orbitals: $d_{xz}$ (0.27 μ$_B$), $d_{yz}$ (0.12 μ$_B$), and $d_{xy}$ (0.11 μ$_B$). The results are the same for B, which again shows the equivalence of A and B. Meanwhile, the major part for C comes from the orbitals $d_{xy}$ (−0.38 μ$_B$) and $d_{z^2}$ (−0.17 μ$_B$). There is also small but non-zero local magnetic moment around the oxygen atoms near the vacancy site: A relatively large spin-down moment is found around the oxygen atom bonded with Pt atom C in the [110] direction (Fig. 6), which is referred as Oc hereafter, with a magnetic moment of −0.11 μ$_B$. When summed over all the atoms, the net magnetic moment of the system approaches zero, which indicates an antiferromagnetic-like coupling between the local magnetic moments.

It is worth stressing that vacancy state **a** and its spin-up counterpart in the valence band contribute differently to the local magnetic moments. The electron in vacancy state **a** has a magnetic moment of −1 μ$_B$ originated from its localized nature, while the valence electrons contribute a net magnetic moment of ~ 1 μ$_B$, which comes not from certain valence states but from all of them, due to the fact that the spin-up valence band contains one more electron than the spin-down valence band. Therefore it will be appropriate to understand the magnetism as antiferromagnetic coupling between the magnetic moment of the unpaired electron in vacancy state **a** (1 $\mu_B$) and the valence electrons (~ 1 $\mu_B$), which yields vanished total magnetic moment. In the right panel of Fig. 3 we can see that a rather flat band is buried in the spin-up valence band at about −0.1 eV. One may expect that the band could give a dominant contribution to the local magnetic moment as a resonant state, but it is found not the case.

Depending on alignment of the local magnetic moments, the excess electrons may have different spin states. The state studied above has zero total magnetic moment and will correspond to the spin singlet state, where the two excess electrons are paired up. We have also considered the other states by restricting the total magnetic moment to be 1 $\mu_B$ or 2 $\mu_B$ (spin triplet state). The total energies of these states are listed in Table III. It is clear that the spin singlet state is the most stable.

The presence of local magnetic moment does not necessarily mean the existence of magnetism, which is a collective phenomenon. To have magnetism, interaction between the vacancies must be sufficiently strong and long-ranged so that the magnetism can sustain at low vacancy concentration and at temperatures of practical interest [15]. The study will be done by analyzing the spatial distribution of the electron density and then by calculating the interaction energy. Figure 7(a) shows the isosurface of electron density of vacancy state **a**, plotted using an isovalue of 0.01 $e/Å^3$. The number of electrons enclosed in the isosurface is ~ 0.55, and the volume enclosed therein is ~ 2.2% of the supercell volume ($V_{cell}$). When enclosed by the isosurface with a larger isovalue of 0.05 $e/Å^3$ (Fig. 5(a)) the number of electrons is reduced to about one-half, ~ 0.26, but the volume is reduce by a factor of ten, ~ 0.28% of $V_{cell}$. This indicates that there is a component in the electron wavefunction delocalized within the supercell although the major part is localized near Vo. Similar feature is found for the other two vacancy states **b** & **c**. The extended component can also be found from the magnetization density by comparing Fig. 6 with Fig. 7(b) which is obtained by using a smaller isovalue. These results suggest that the magnetic interaction within the supercell can be sufficiently long-ranged. Similar results are reported in the semiconducting materials GaN and BN with the presence of native defects [15].

We go on to estimate the strength of magnetic coupling by introducing two O vacancies into a (2×2×3) supercell of β-$PtO_2$. Considering the ratio between Pt and O atoms, the system with two O vacancies in a (2×2×3) supercell is labeled as $PtO_{1.917}$. The magnetic state with zero total magnetic moment is found to be the most stable, which is similar to the system with one O vacancy. Compared to spin-unpolarized or non-magnetic (NM) case, the total energy of the spin-polarized configuration (AFM coupled) with one O vacancy ($PtO_{1.958}$) is lowered by ~ 0.25 eV in a (2×2×3) supercell and by ~ 0.24 eV in a (3×3×4) supercell. The small difference indicates, the vacancy-vacancy interaction with the periodic image is negligible when using a (2×2×3) or larger supercell for simulation. Consequently, the strength of magnetic coupling between the two Vo sites can be estimated as follows:

$$E_{int} = (E_{NM} - E_{AFM}) - 2\Delta E, \qquad (3)$$

where $E_{NM}$ and $E_{AFM}$ is the total energy of spin-unpolarized (NM) and spin-polarized configuration (AFM) of PtO$_{1.917}$, respectively. The quantity $\Delta E = 0.24$ eV, is the energy gain due to the coupling between the local magnetic moments that induced by an isolated O vacancy. The magnetic coupling strength $E_{int}$ is calculated to be ~ 0.11 eV when the distance between two Vo sites is 4.0 Å, and the value is reduced to ~ 0.05 eV when the Vo-Vo distance is enlarged to 7.2 Å. Such coupling strength suggests that the magnetism can survive even at room temperature at the vacancy concentration where the Vo-Vo distance is ~ 7 Å on average.

Figure 8 shows the electron density difference after the creation of an oxygen vacancy. The quantity $\Delta\rho$ in Eq. (2) is determined by three factors: 1) Charge transfer due to the removal of one O atom; 2) Spin polarization of the excess electrons; 3) Redistribution of charge density due to structural relaxation, including atomic position mismatch with comparison to perfect system. The third factor is a minor part and somewhat artificial, because the structural relaxation is small and the contribution to charge difference can be largely reduced when the atomic positions of the remaining part are the same as defect-free system. Therefore, we focus on the first two factors: charge transfer and spin polarization. To get $\rho$[PtO$_{1.958}$], calculation was done on the PtO$_{1.958}$ system with all the atomic positions kept the same as in defect-free β-PtO$_2$. For spin-unpolarized $\rho$[PtO$_{1.958}$], the quantity $\Delta\rho$ calculated using the triplet state is shown in Fig. 8(a), and that calculated using the singlet state is shown in Fig. 8(b). For spin-polarized $\rho$[PtO$_{1.958}$], the $\Delta\rho$ calculated using the triplet and singlet O is shown in Fig. 8(c) and Fig. 8(d), respectively. The common feature is that the three Pt atoms A, B, and C gain electrons while the Vo site losses electrons when one oxygen atom is removed.

Regarding the different spin state of O atom, the electron density difference near Vo site has much smaller value in Figs. 8(b) and 8(d) than in Figs. 8(a) and 8(c). In Figs. 8(a) and 8(c), besides electron loss, there is also electron gain within a small region near the Vo site. This is due to the fact that the O atom itself is spin singlet state (zero net magnetic moment) in the defect-free β-$PtO_2$ (non-magnetic) while it is spin triplet state in the gas phase, which results in significant difference in the electron distribution. On the contrary, if the electron density of spin singlet state is employed in the calculation, no electron gain is seen at the Vo site and one sees only electron loss contours with nearly spherically symmetric distribution (Figs. 8(b) and 8(d)). This is consistent with the picture that an O atom will leave excess electrons behind the remaining system when it is removed from bulk oxides. For calculation using the electron density of fully relaxed structure of $PtO_{1.958}$, the major results at Pt A, B, C and the Vo site are similar, except for some modified features at the O atoms nearby due to the displaced atomic position with reference to defect-free system.

The presence of excess electrons will induce the reorganization of spin-up and spin-down electron occupancy within each $5d$ orbital of Pt, due to the strong on-site Coulomb repulsion. This is important and will give rise to the possibility of appearance of local magnetic moment. When reducing the values of $U_{eff}$ to test the effects of on-site Coulomb repulsion, we find that, for $U_{eff} < 5.5$ eV, the spin moments of all the atoms are negligible (≤ 0.1 $\mu_B$). This reveals the key role of on-site Coulomb repulsion within the 5*d* shell of Pt.

Compared to some commonly studied strongly-correlated materials [13, 62-64], whose $U_{eff}$ usually ranges from ~ 5 eV to ~ 6 eV, the value of $U_{eff}$ = 7.5 eV is somewhat large. Nevertheless, this value still locates in a reasonable range [65]. In fact, we have also found systems with a theoretical value of U = 8 eV (ZnO, ZnS) [66, 67] or even a higher value of 9 eV (PtO) [58]. Moreover, as mentioned above, our GW calculation gives strong support for choosing such a value.

### B. Charged systems

Now we analyze the charge state of the oxygen-deficient system. In the above analysis, we assumed a charge-neutral state (denoted as $V_O^0$). Here we make comparison for the stability among differently charged systems. To reduce the long-range Coulomb repulsion between the neighboring charged images under periodic boundary condition, a (3×3×4) supercell is employed in our calculation, and the system with one O vacancy is denoted as PtO$_{1.986}$. We use the symbol δ*q* to denote net charge of the system. Three charge states are considered here: δ*q* = −1*e* ($V_O^-$), 0 ($V_O^0$), and +1*e* ($V_O^+$) with *e* the elementary electric charge. The $V_O^+$ ($V_O^-$) state was simulated by subtracting (adding) one electron from (to) the PtO$_{1.986}$ system. For the charged systems, a homogeneous background charge with opposite sign was introduced so that the total energy calculation is converged [68]. Since the total number of electrons is not conserved, the quantity considered for comparison is the grand potential [69]:

$$\Omega = E - TS - \mu N = E - TS - \mu(N_0 + \Delta N),  \quad (4)$$

where $(E - TS)$ is the free energy term, $\mu$ is the chemical potential of electrons, $N_0$ is the number of electrons in the charge-neutral system, and $\Delta N$ ($= -\delta q/e$) is the change in the total number of electrons. The entropy $S$ is mainly contributed from lattice vibration of atoms. Compared to defect-free structure, the maximum lattice distortion of the three charge states is less than 0.5% because of low vacancy concentration. Near the Vo site, the maximum bond length and bond angle difference is less than 1.5% and 1%, respectively. Thus the difference of entropy $S$ at different charge state of $PtO_{1.986}$ is negligible, especially at low temperature. The quantity we need to compare is reduced to

$$\tilde{\Omega} = E - \mu \Delta N, \quad (4')$$

The grand potentials for $V_O^-$ ($\Delta N = +1$), $V_O^0$ ($\Delta N = 0$), and $V_O^+$ ($\Delta N = -1$) are as follows:

$$\tilde{\Omega}[V_O^-] = E[V_O^-] - (\mu - \varepsilon_{VBT}) - \varepsilon_{VBT}; \quad \tilde{\Omega}[V_O^0] = E[V_O^0]; \text{ and } \tilde{\Omega}[V_O^+] = E[V_O^+] + (\mu - \varepsilon_{VBT}) + \varepsilon_{VBT},$$

where $\varepsilon_{VBT}$ is the top of spin-up valence band at charge-neutral state. The results are shown in Fig. 9, for $\tilde{\Omega}$ (referenced to $\tilde{\Omega}[V_O^0]$) as a function of $(\mu - \varepsilon_{VBT})$. It is clear that there exists a critical point ($\mu - \varepsilon_{VBT} \sim 0.025$ eV), at which the three charge states are of the same stability. When the crossing point is magnified (see inset of Fig. 9), one sees that the charge-neutral system ($V_O^0$) is actually the most stable when 0.011 eV $\leq (\mu - \varepsilon_{VBT}) \leq$ 0.039 eV. Below and above this region, the charge states $V_O^+$ and $V_O^-$ are energetically favored, respectively. When the chemical potential $(\mu - \varepsilon_{VBT})$ of charge systems $V_O^0$ and $V_O^+$ locates in the region [0.011 eV, 0.025 eV], while $V_O^0$ and $V_O^-$ in the region [0.025 eV, 0.039 eV], the energy difference of each two charge states is no higher than 16 meV, which is within the error bar of computational accuracy. The same result is obtained when using higher energy cut-off (800 eV) of plane waves and/or denser k-mesh (3×3×3) for total energy calculation. Therefore, each group of charge states can be regarded as having the same stability, or the same probability to be observed in experiment. Beyond this region, the charge state $V_O^-$ is stable when $0.039 \text{ eV} \leq (\mu - \varepsilon_{VBT}) \leq 1.2$ eV, and $V_O^+$ is stable when $0.00 \text{ eV} \leq (\mu - \varepsilon_{VBT}) \leq 0.011$ eV. The following analysis will focus on the nonzero charged states.

We have examined the electronic density of states (DOS) of charged systems $V_O^\pm$, and found the overall features are slightly modified comparing to the charge-neutral one. Our studies on a series of negatively charged systems ($-1e \leq \delta q < 0$) show that, the gap between the top of spin-up valence band and the vacancy state **b** (Fig. 3) decreases continuously from 90 meV ($\delta q \sim 0$) to ~ 15 meV ($\delta q = -1e$) when the vacancy state **b** is gradually filled up by one electron. For $\delta q = -1e$, the most notable change in DOS is that vacancy state **c** is merged into spin-down conduction band. As for $0 \leq \delta q < +1e$, the DOS features are the almost same as charge-neutral state ($\delta q = 0$), except that the Fermi level shifts down to the top of valance band and the nearby spin-down vacancy state (peak **a** in Fig. 3) is left empty. The band gap is kept at ~ 1.2 eV in all cases.

In addition, our calculation demonstrates that, net total magnetic moment will appear when subtracting (adding) one more electron from (into) the system. The total magnetic moment is calculated to be 1 $\mu_B$ for $V_O^{\pm}$. For the negatively charged system ($V_O^-$), we have also considered the other spin multiplicity states with a total magnetic moment ($M_{total}$) of 0, 2, and 3 $\mu_B$, and found that the case $M_{total} = 1$ $\mu_B$ stands out to be the most stable. This implies that, for $V_O^+$ where no electrons are in the vacancy state, the total magnetic moment is solely the sum of spin magnetic moment of the valence electrons, which is ~ 1 $\mu_B$. The result for $V_O^-$ can be similarly explained if the two electrons in the vacancy states are paired up and give no net contribution to total magnetic moment. These results indicate, the magnetic behavior can be explained within the framework of rigid band model, which assumes that the electronic states in the DOS diagram are kept unchanged when charging up the system by shifting the position of Fermi level.

To support the above explanation, we investigated the electron distribution. The three charge states studied here are directly related to the filling of vacancy states **a** & **b**: Both **a** & **b** are empty for $\delta q = +1e$, while **a** is filled up and **b** is left empty for $\delta q = 0$, and both **a** & **b** are filled up for $\delta q = -1e$. Thus, if the rigid band model is applicable, the charge density of each vacancy state should remain unchanged under different charge states. Specially, the relation $\rho_{\mathbf{a}}[V_O^+] = \rho_{\mathbf{a}}[V_O^0]$ and $\rho_{\mathbf{b}}[V_O^-] = \rho_{\mathbf{b}}[V_O^0]$ should hold for vacancy states **a** & **b.**

The electron density of vacancy states **a** ($\rho_{\mathbf{a}}[V_O^0]$) and **b** ($\rho_{\mathbf{b}}[V_O^0]$) at charge-neutral state of the PtO$_{1.986}$ system are shown in Fig. 10. The spatial distribution of both $\rho_{\mathbf{a}}[V_O^0]$ and $\rho_\mathbf{b}[V_O^0]$ is the nearly same as the corresponding $\rho_\mathbf{a}$ and $\rho_\mathbf{b}$ of PtO$_{1.958}$ in the smaller (2×2×3) supercell. By examining the corresponding quantities of singly-charged states $\rho_\mathbf{a}[V_O^+]$ and $\rho_\mathbf{b}[V_O^-]$, we find only small difference between their charge-neutral counterpart. That means, $\rho_\mathbf{a}[V_O^+] \approx \rho_\mathbf{a}[V_O^0]$ and $\rho_\mathbf{b}[V_O^-] \approx \rho_\mathbf{b}[V_O^0]$. Thus, the rigid band model is a good approximation for explaining the appearance of net magnetic moment in the singly-charged PtO$_{1.986}$ system. The small difference at different charge state reflects the perturbation on the total charge density upon adding or removing one electron into/from the charge-neutral system.

The isosurface and 2D contours of magnetization density for charge states $V_O^\pm$ are displayed in Fig. 11. We provide detailed description by analyzing the electron density as done above for the charge-neutral state. For $V_O^-$, integrals inside a sphere shell (radius = 1.3 Å) give the total number of spin-up (spin-down) electrons for Pt atoms A, B, and C to be 4.61 (3.91), 4.61 (3.92), and 3.87 (4.56), respectively. The local magnetic moment for Pt A, B, and C is calculated to be 0.70 μ$_B$, 0.69 μ$_B$, and −0.69 μ$_B$, respectively. As shown in Figs. 11(a)-(b), along each Pt-Vo direction, considerable magnetization appears around the three O atoms bonded with Pt A, B, and C, with the corresponding magnetic moment: 0.l5 μ$_B$, 0.15 μ$_B$, and −0.12 μ$_B$. Compared to charge-neutral state, the local magnetic moments are enlarged, and the major part of the added electron contributes to the spin-up component.

Similarly, for $V_O^+$, the number of spin-up (spin-down) electrons of Pt A, B, and C is counted to be 4.56 (3.99), 4.56 (4.00), and 3.90 (4.53). The local magnetic moment is 0.57 $\mu_B$, 0.56 $\mu_B$, and −0.63 $\mu_B$ for Pt A, B, and C, respectively. The total number of electrons is almost the same around A, B, and C when comparing to charge-neutral system. Compared to $V_O^0$ and $V_O^-$ state, a notable change is in direction of the local magnetic moment at the Oc site (Figs. 11(c)-(d)), with a value of ~ 0.03 $\mu_B$. This is due to the removal of one spin-down electron at vacancy state **a**, which is localized at the Oc site and other O atoms surrounding Pt C (Fig. 5(a)). Variation of spin-up and spin-down electrons results in the change of local magnetic moment direction.

### C. Discussion

In the above we found that vacancy states **a** and **b** (which are occupied and unoccupied, respectively, when charge-neutral) are located quite close to the valence band top. In spite of this fact, their localization length is not long enough to affect band dispersion even in the smaller (2×2×3) supercell. This is contrary to the shallow states in typical semiconductors like Si or GaAs. The electrical conduction, or the carrier supply, via the vacancy state is therefore expected to be slow when the defect concentration is dilute. However, when more vacancies are introduced, metallic conduction is possible. Such results have been reported in the amorphous films of oxygen-deficient $PtO_2$ [53]. Another aspect we would like to point out is the small structural relaxation of this oxygen-deficient β-$PtO_2$, for both charge-neutral and singly-charged systems. This is distinguished from the other systems. Usually oxides have large lattice relaxation upon vacancy formation and/or charging [70, 71].

Although quantitatively relating the electrocatalytic activity with the oxide layer is beyond the scope of the present paper, our calculation for the bulk property provides an aspect of the catalytic property. When the oxide layer covers the surface, the reaction rate will be significantly lowered by the insulating property of β-PtO$_2$. When defects are introduced, however, they can provide a conducting channel for the carrier. Our calculation suggests that the oxygen vacancy is a candidate for such defect in the *p*-type condition. When, in addition, the defect level is aligned to the affinity level (ionization potential) of the reactant, the vacancy can play a role as an electron donor (acceptor). In that case, because of the equivalent stability of V$^0$ and V$^\pm$, the vacancy can act as a donor as well as an acceptor at the same time. The 2*p* character of the vacancy state implies that the affinity level or the ionization level should be aligned to it. In the context of the oxygen reduction reaction (ORR), the level alignment to the reaction intermediates, O$_2$, O, OH, and OOH, is required. Among them, donating an electron to the adsorbed O$_2$ is particularly important because breaking O-O bond is considered as the rate-limiting step on Pt and the donated electron is considered to occupy the anti-bonding state [72]. It is therefore a target of future study to investigate the level alignment using, for example, a recently proposed computational scheme of Sprik group [73]. When the oxide film of only a few atomic layer is formed, the electrons will be able to tunnel through the layer. It is possible that the layer may not significantly reduce the catalytic activity as recently found for MgO thin layer formed on Ag [74].

It is noteworthy to point out here that, in a previous work [75], the spin state of an O$_2$ molecule plays a key role in its dissociation process on Al(111). Therefore, the presence of magnetism in oxygen-deficient β-PtO$_2$ may also have nontrivial effect on the dissociation of O$_2$ on β-PtO$_2$ covered Pt electrode via magnetic interactions.

Charge transfer associated with removal of one O atom is expected as an origin of the appearance of local magnetic moment. The electrons transferred to neighboring Pt atoms of Vo site cause reorganization of 5$d$ occupancy because of the strong on-site Coulomb repulsion, and local magnetic moment is induced by the difference of spin-polarized electrons. Although the GGA+U scheme has been verified by our GW calculation, which is more accurate, the prediction regarding magnetism needs to be further tested by experimental measurement like the electron spin resonance (ESR), which in turn can be used to access the strength of on-site Coulomb repulsion.

**IV. Conclusion**

To summarize, the properties of oxygen vacancy in β-PtO$_2$ have been studied using first-principles calculations within the DFT-GGA+U scheme. With this method, the structural parameters and band gap of crystalline β-PtO$_2$ are reasonably reproduced. The validity of the value of parameter U employed in GGA+U calculation is further justified by GW calculation. The ground state is spin-polarized when an oxygen vacancy is present. Local magnetic moment appears majorly at the three nearest neighboring Pt atoms of the vacancy site. The total magnetic moment is zero for charge-neutral state ($V_O^0$), while it is 1 $\mu_B$ for singly-charged states $V_O^+$ and $V_O^-$. The magnetism can be understood from antiferromagnetic coupling between the unpaired electrons in the vacancy state and the itinerant valence electrons. The local magnetic moment of Pt atoms originates from the occupation reorganization of 5$d$ electrons, due to the Coulomb interaction upon electrons transferred from the vacancy site. The magnetic coupling is found to be long-ranged and is expected to survive even at room temperature. When the chemical potential µ varies from 0.011 eV to 0.039 eV (referenced to valence top), the $V_O^0$ state is the most stable. In this region, the difference of grand potential of the three charge states is very small and can be regarded as the same within the computational accuracy. Therefore, this region can be regarded as a triple point and large fluctuations between different charge and magnetic states are expected. Beyond this region, singly-charged states ($V_O^\pm$) are stable with net magnetic moment of ~ 1 $\mu_B$. In particular, the negatively charged state ($V_O^-$) is the most stable when $0.039 \leq \mu \leq 1.2$ eV with reference to valence top. So, the oxygen-deficient β-PtO$_2$ is nearly a negative-U system. Within the rigid band model, the filling of vacancy states accounts for the variation of magnetic state. The vacancy states studied here are close to the valence top and can serve as an electron donor and acceptor under $p$-type condition and may play an important role in catalysis, in particular, the oxygen reduction reaction.

**Acknowledgement**

This work is supported by the Global Research Center for Environment and Energy based on Nanomaterials Science (GREEN) at National Institute for Materials Science (NIMS). The first-principles calculations were carried out by the supercomputer (SGI Altix) of NIMS. The 2D contours and 3D isosurfaces are drawn using the program XCrySDen [76-78]. One of the authors (Y. Y.) would like to thank Professor Ding-Sheng Wang of CAS for his reading and helpful comments on the manuscript.

**Table I.** Cell parameters of β-PtO$_2$ obtained by DFT-GGA and GGA+U calculations, with comparison to experimental data.

| Method | $a$ (Å) | $b$ (Å) | $c$ (Å) | $\Delta V/V_{expt}$ (%) |
|---|---|---|---|---|
| GGA | 4.611 | 4.557 | 3.191 | 5.06 |
| GGA+U | 4.535 | 4.577 | 3.130 | 1.80 |
| [a]Expt | 4.4839 | 4.5385 | 3.1360 | --- |

[a] Reference [42].

**Table II.** Calculated *lm*-component of the vacancy states **a**, **b**, and **c** (see Fig. 3).

| State/lm | $s$ | $p_y$ | $p_z$ | $p_x$ | $d_{xy}$ | $d_{yz}$ | $d_{z^2}$ | $d_{xz}$ | $d_{x^2-y^2}$ |
|---|---|---|---|---|---|---|---|---|---|
| **a** | 0.014 | 0.463 | 0.028 | 0.236 | 0.042 | 0.044 | 0.016 | 0.042 | 0.045 |
| **b** | 0.018 | 0.198 | 0.170 | 0.276 | 0.043 | 0.089 | 0.033 | 0.077 | 0.027 |
| **c** | 0.063 | 0.061 | 0.095 | 0.097 | 0.150 | 0.115 | 0.054 | 0.185 | 0.011 |

**Table III.** Total and local magnetic moment (Pt A, B, C and the Oc site in Fig. 6) and relative energy to ground state for different magnetic states of PtO$_{1.958}$.

| $M_{total}$ (μ$_B$) | $M_A$ (μ$_B$) | $M_B$ (μ$_B$) | $M_C$ (μ$_B$) | $M_{Oc}$ (μ$_B$) | $\Delta E$ (eV/cell) |
|---|---|---|---|---|---|
| 0 | 0.59 | 0.58 | -0.67 | -0.11 | 0.00 |
| 1 | 0.64 | 0.65 | -0.67 | -0.02 | 0.05 |
| 2 | 0.61 | 0.62 | 0.70 | 0.13 | 0.14 |

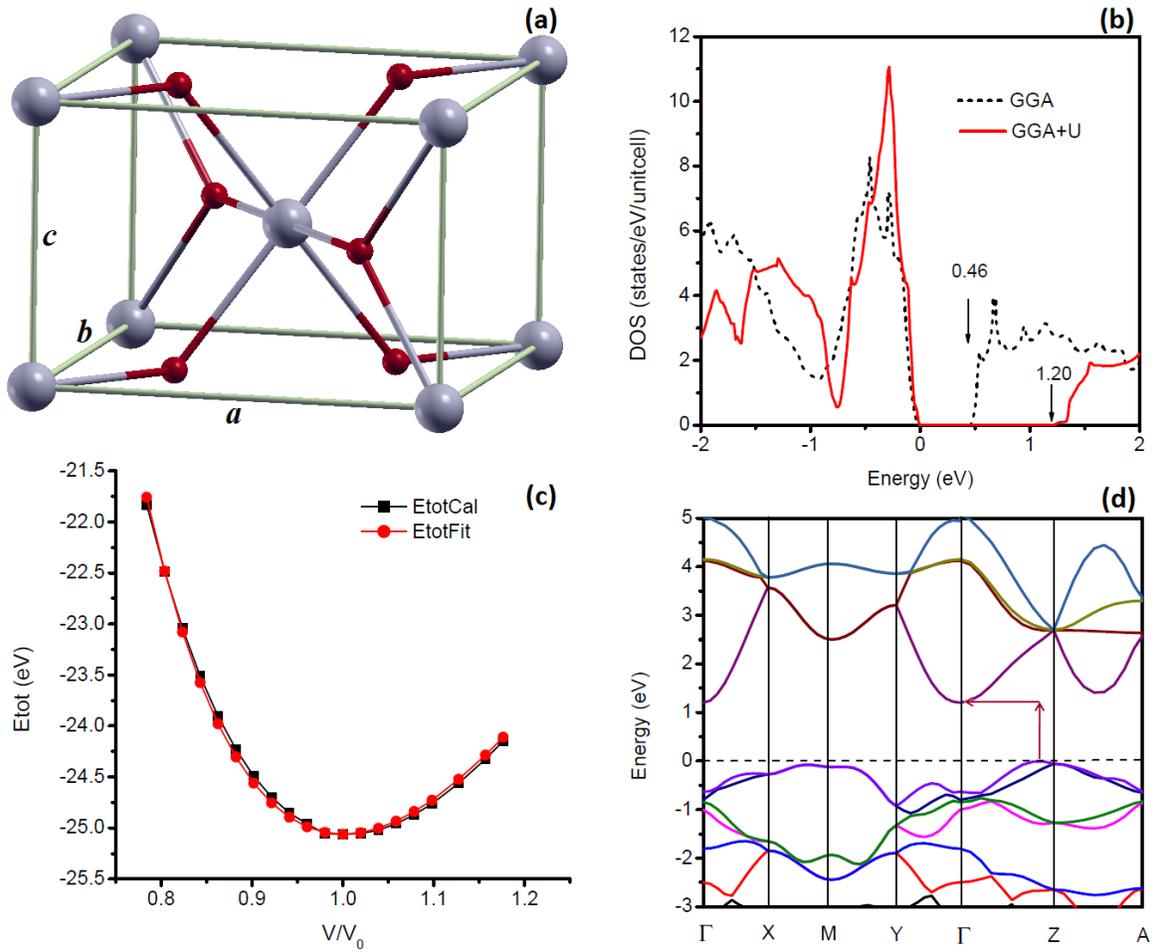

**FIG. 1** (color online) Crystal and electronic properties of β-PtO$_2$: (a) Unit cell, with Pt atoms represented by large (gray) balls, and O atoms by small (red) balls. (b) Calculated total DOS (unit: states/eV/cell) with GGA (black dashed line) and GGA+U (red solid line). (c) Total energy as a function of normalized volume V/V$_0$, V$_0$ is the equilibrium volume. (d) Calculated energy band structure with GGA+U. The Fermi level (highest occupied level) is set at zero. The indirect band gap is indicated by two arrows. The direct coordinates of the k-points in Brillouin zone: Γ = (0, 0, 0), X = (0.5, 0, 0), M = (0.5, 0.5, 0), Y = (0, 0.5, 0), Z = (0, 0, 0.5), A = (0.5, 0.5, 0.5). The convention applies to all the figures.

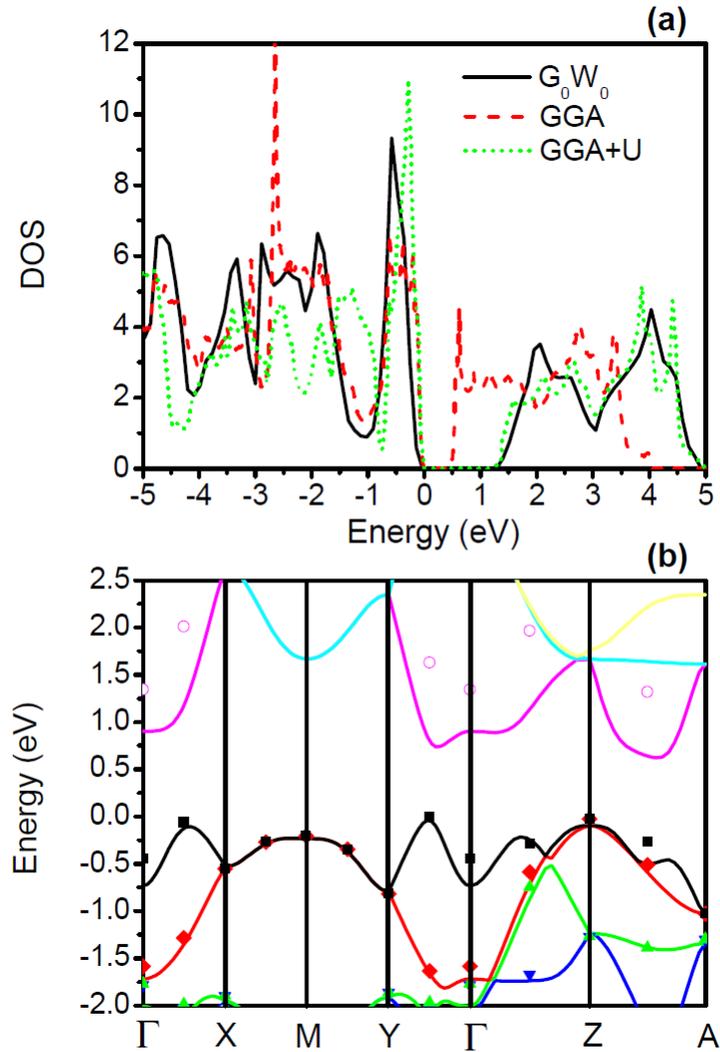

**FIG. 2** (color online) (a) Calculated electronic density of states (DOS) of β-PtO$_2$ by using GGA, GGA+U and G$_0$W$_0$ method. (b) Energy bands of β-PtO$_2$, calculated using GGA (solid lines) and G$_0$W$_0$ method (scattered dots) along some lines joining the high-symmetry points in the k-space.

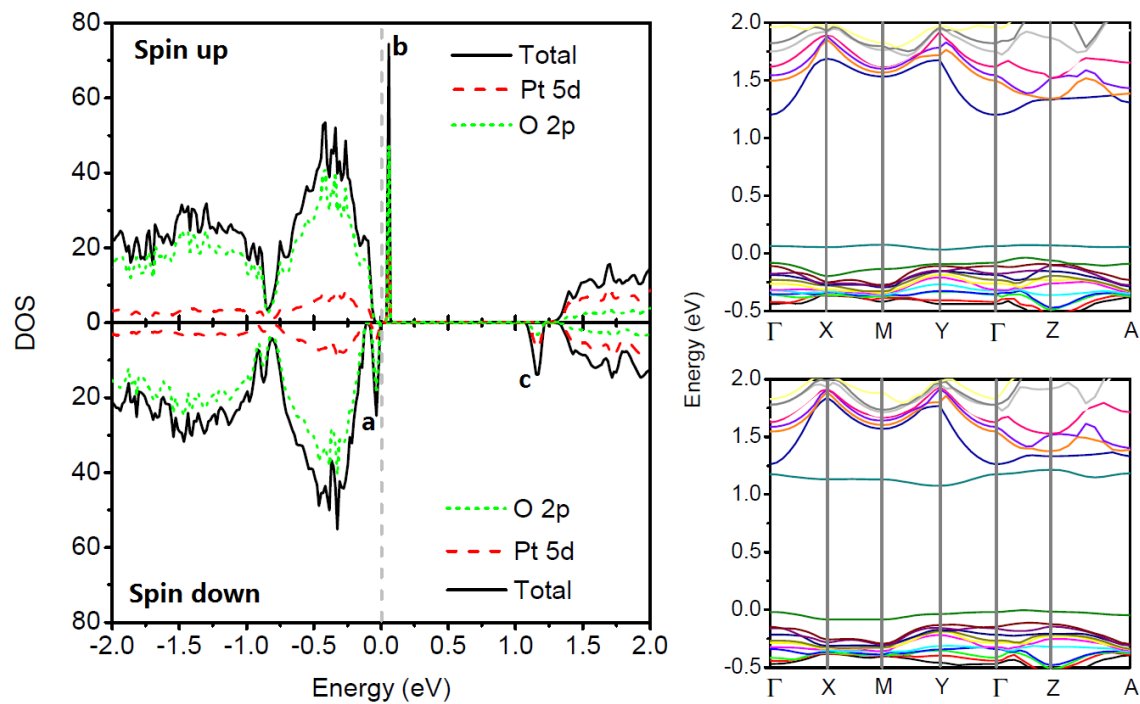

**FIG. 3** (color online) Left panels: Spin-polarized DOS (unit: states/eV/supercell) for PtO$_{1.958}$. The vacancy states are labeled by **a**, **b**, and **c**. The Fermi level (set at 0 eV for all panels) is indicated by a gray dashed line. Right panels: Calculated band structure of spin-up (upper panel) and spin-down (lower panel) electronic states along some high symmetry k-point lines of BZ. The unit of DOS applies to all the other figures.

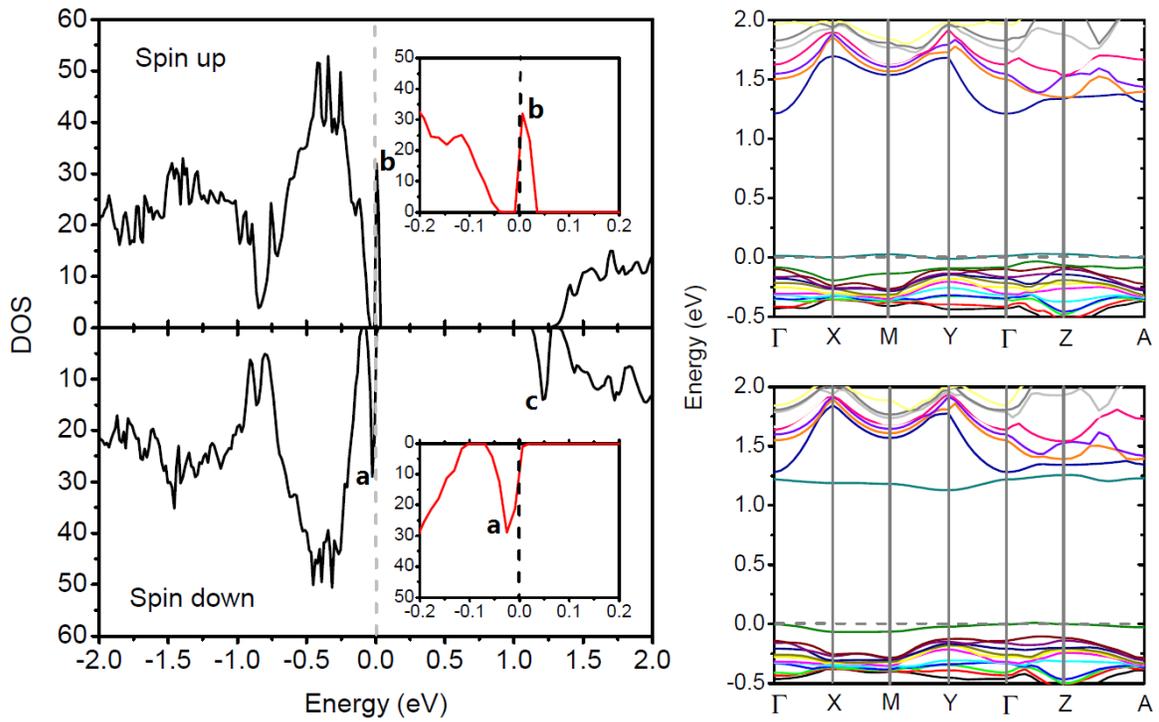

**FIG. 4** (color online) Same as Fig. 3, but for a slightly modified $PtO_{1.958}$ structure with slightly higher total energy (by ~ 9 meV). The crossing region between vacancy states **a** & **b** and the Fermi level (dashed line) is highlighted in the inset.

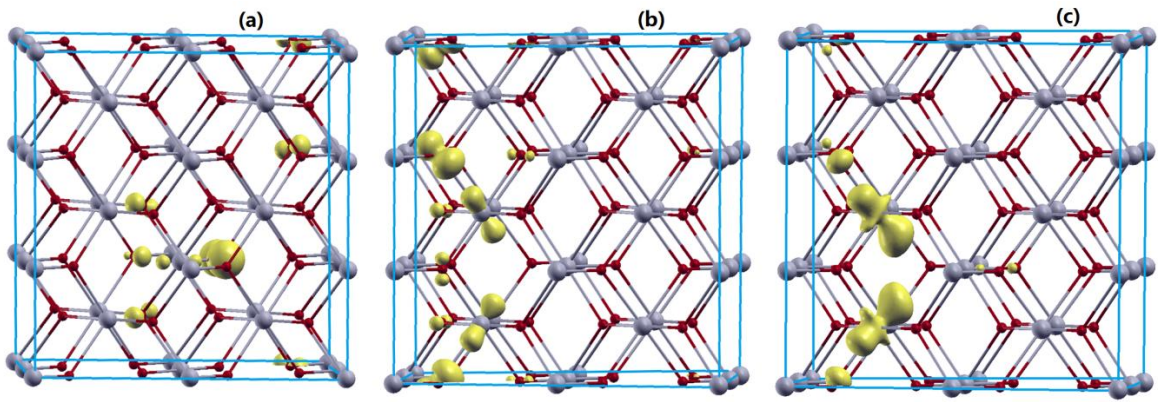

**FIG. 5** (color online) Panels (a)-(c), from left to right: Isosurface of electron density of vacancy states **a**, **b** and **c**, respectively. The isovalue is 0.05 $e/\text{Å}^3$.

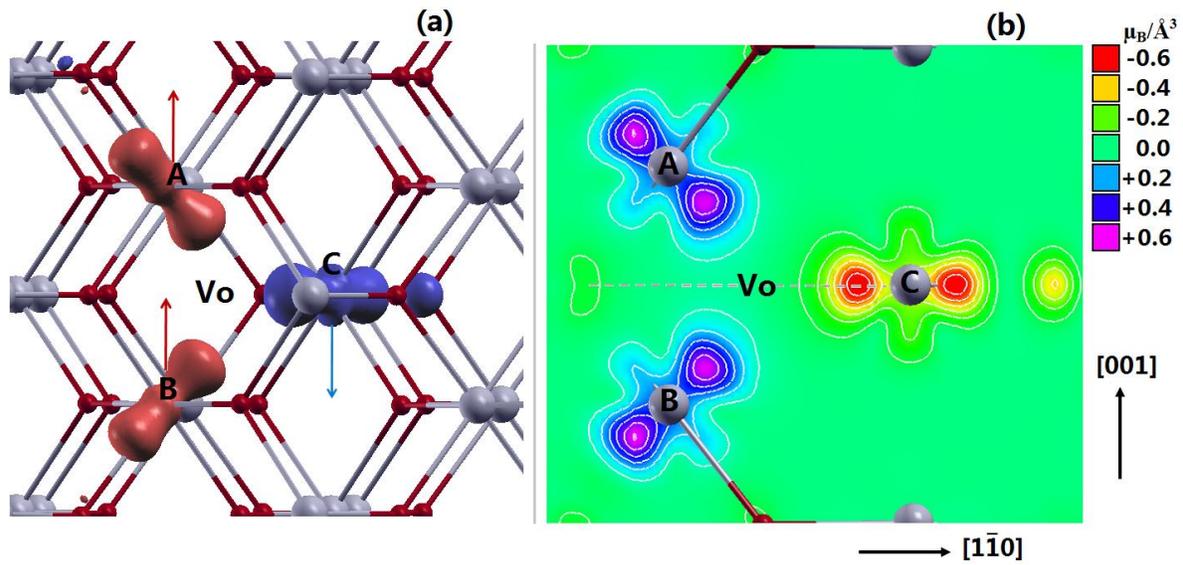

**FIG. 6** (color online) (a) Magnetization density of $PtO_{1.958}$. The isovalue for spin-up (red) and spin-down (blue) isosurface is 0.2 $\mu_B/\text{Å}^3$ and -0.2 $\mu_B/\text{Å}^3$, respectively. (b) Two-dimensional contours in the plane cutting through the three Pt atoms and the Vo site.

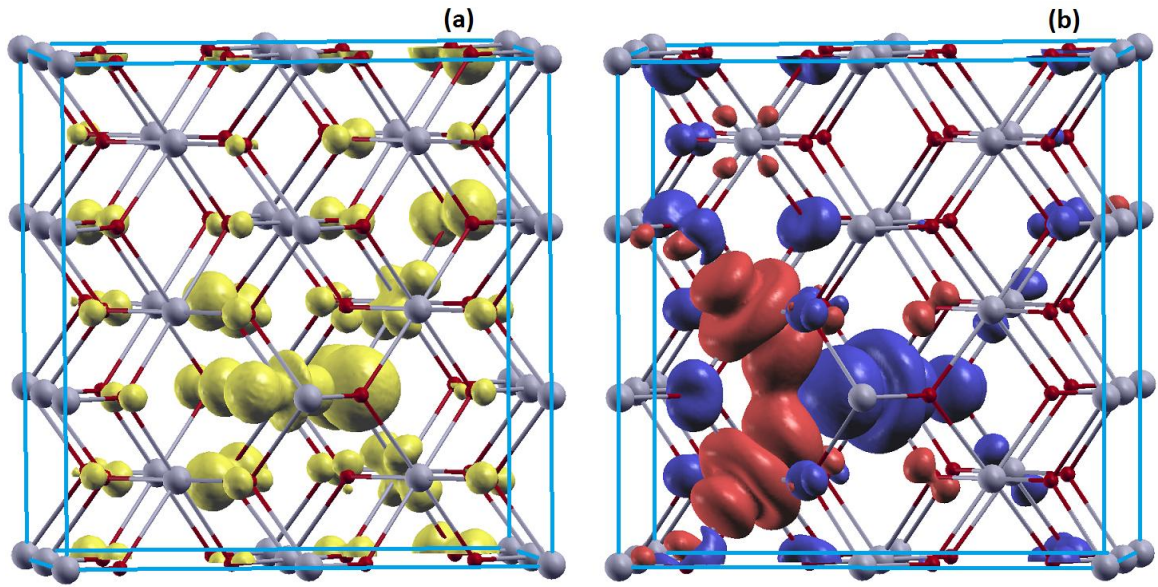

**FIG. 7** (color online) (a) Isosurface of electron density of vacancy state **a**, plotted using an isovalue of 0.01 $e$/Å$^3$. (b) Isosurface of magnetization density of PtO$_{1.958}$, plotted using an isovalue of ±0.02 $\mu_B$/Å$^3$: Red for spin-up and blue for spin-down.

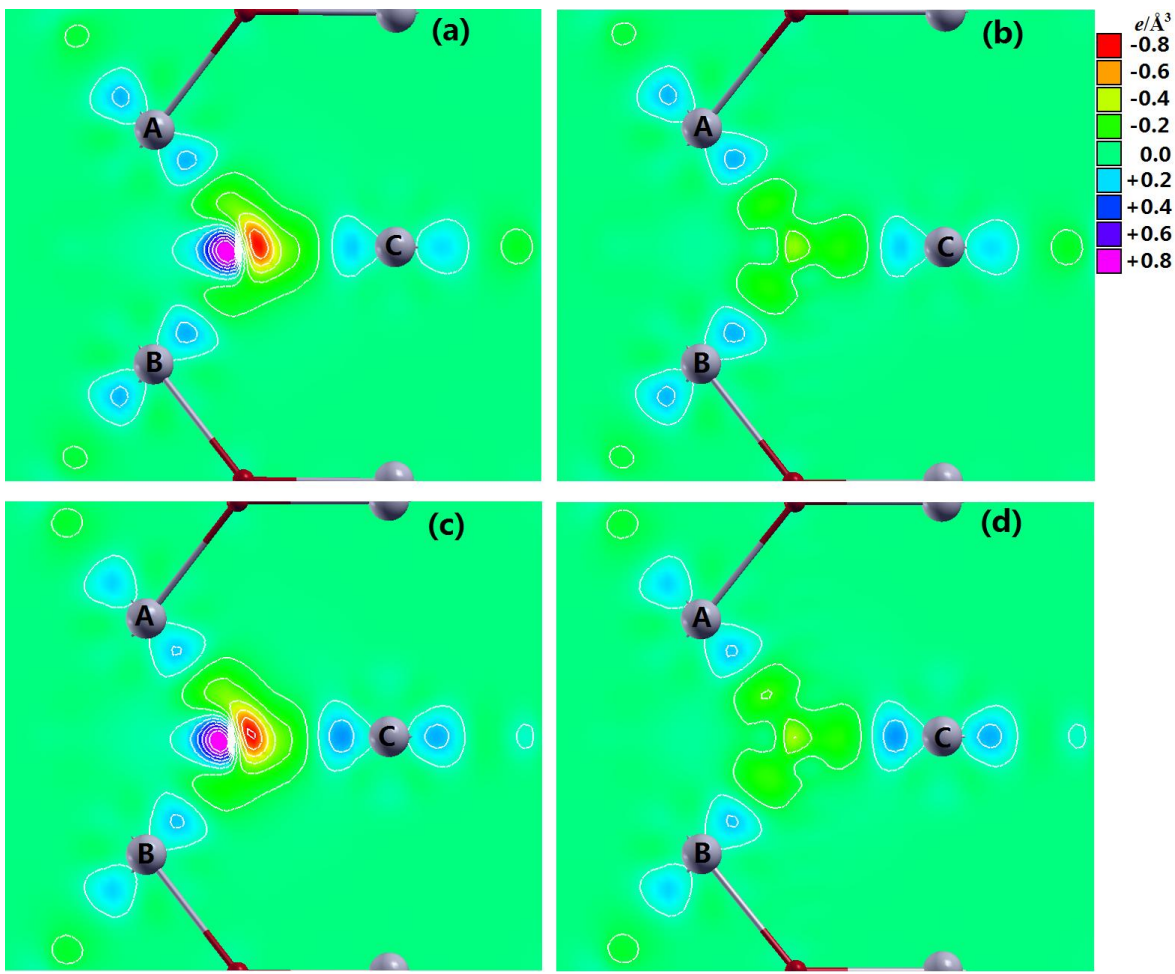

**FIG. 8** (color online) Two-dimensional contours of electron density difference around the vacancy site of $PtO_{1.958}$. The electron densities employed for calculation of different spin polarization states are categorized in the following panels: (a) Spin-unpolarized $PtO_{1.958}$, spin triplet O; (b) Spin-unpolarized $PtO_{1.958}$, spin singlet O; (c) Spin-polarized $PtO_{1.958}$, spin triplet O; (d) Spin-polarized $PtO_{1.958}$, spin singlet O. In all case the O atom is in gas phase. The Pt atoms are marked by A, B, C. The vacancy locates at the center of mass of the ABC triangle.

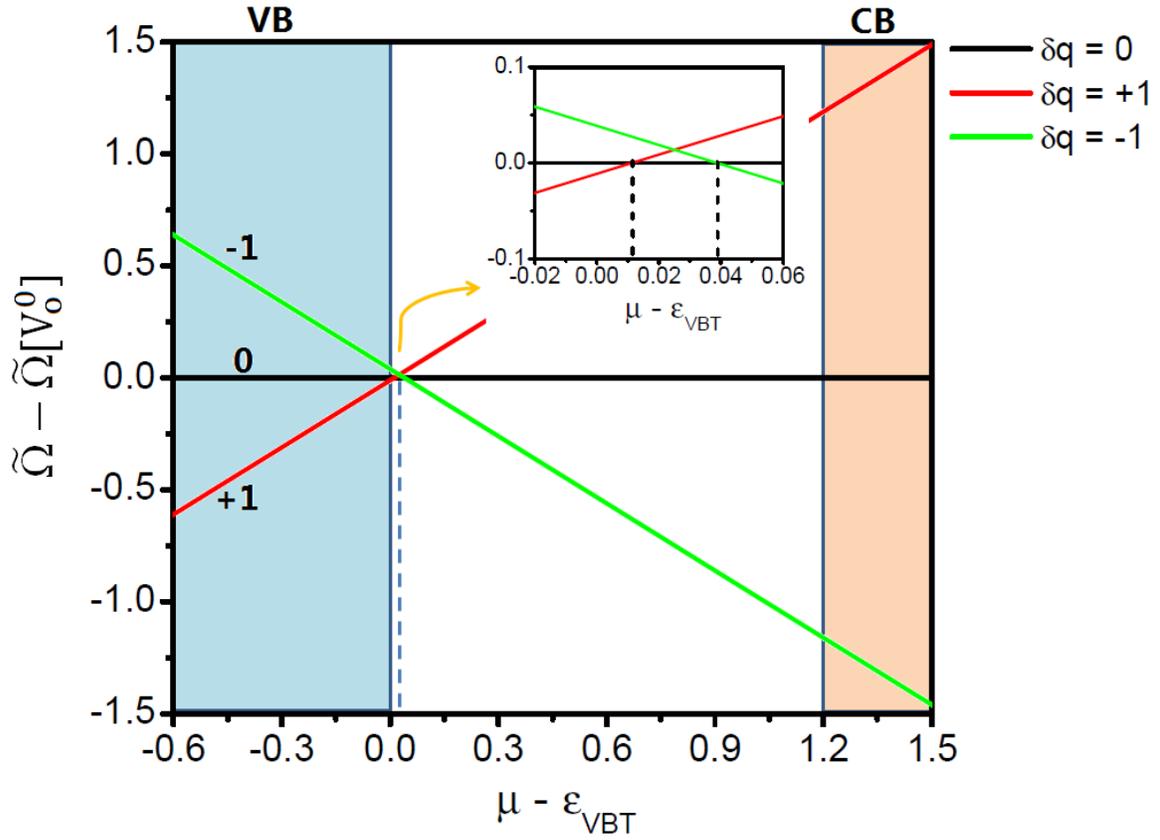

**FIG. 9** (color online) Calculated grand potential for various charge states of the PtO$_{1.986}$ system: $\delta q = 0$, $-1e$, and $+1e$. The unit for energy is eV. The valence band (VB) and conduction band (CB) are illustrated by the shading regions. The crossover region of the three charge states is highlighted in the inset.

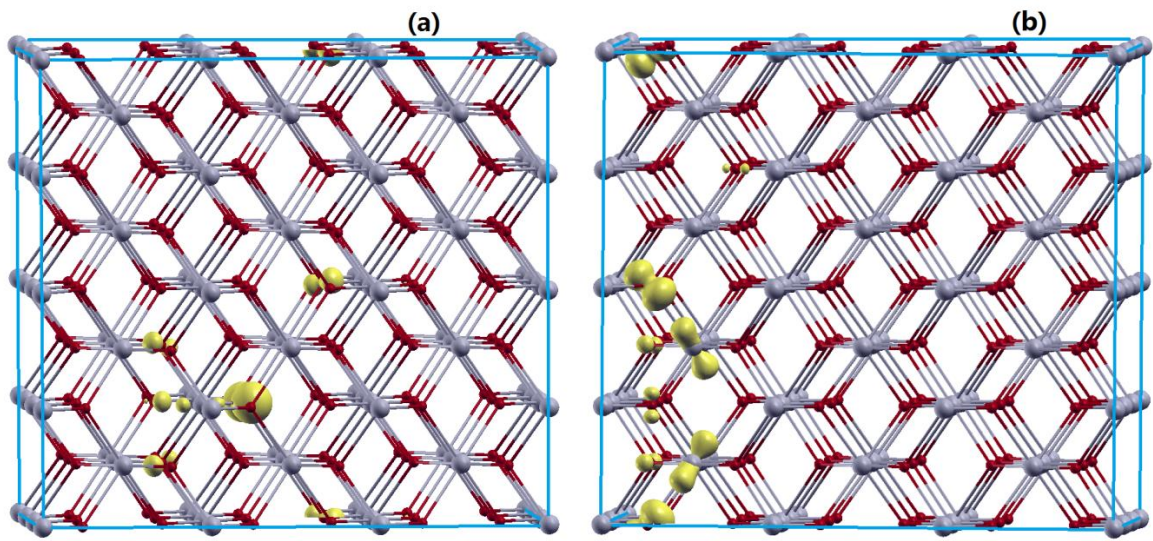

**FIG. 10** (color online) Isosurface of electron density of vacancy states **a** & **b** in the charge-neutral PtO$_{1.986}$ system ($V_O^0$): $\rho_\mathbf{a}[V_O^0]$ (panel (a)) and $\rho_\mathbf{b}[V_O^0]$ (panel (b)). The isovalue of density is 0.05 $e$/Å$^3$.

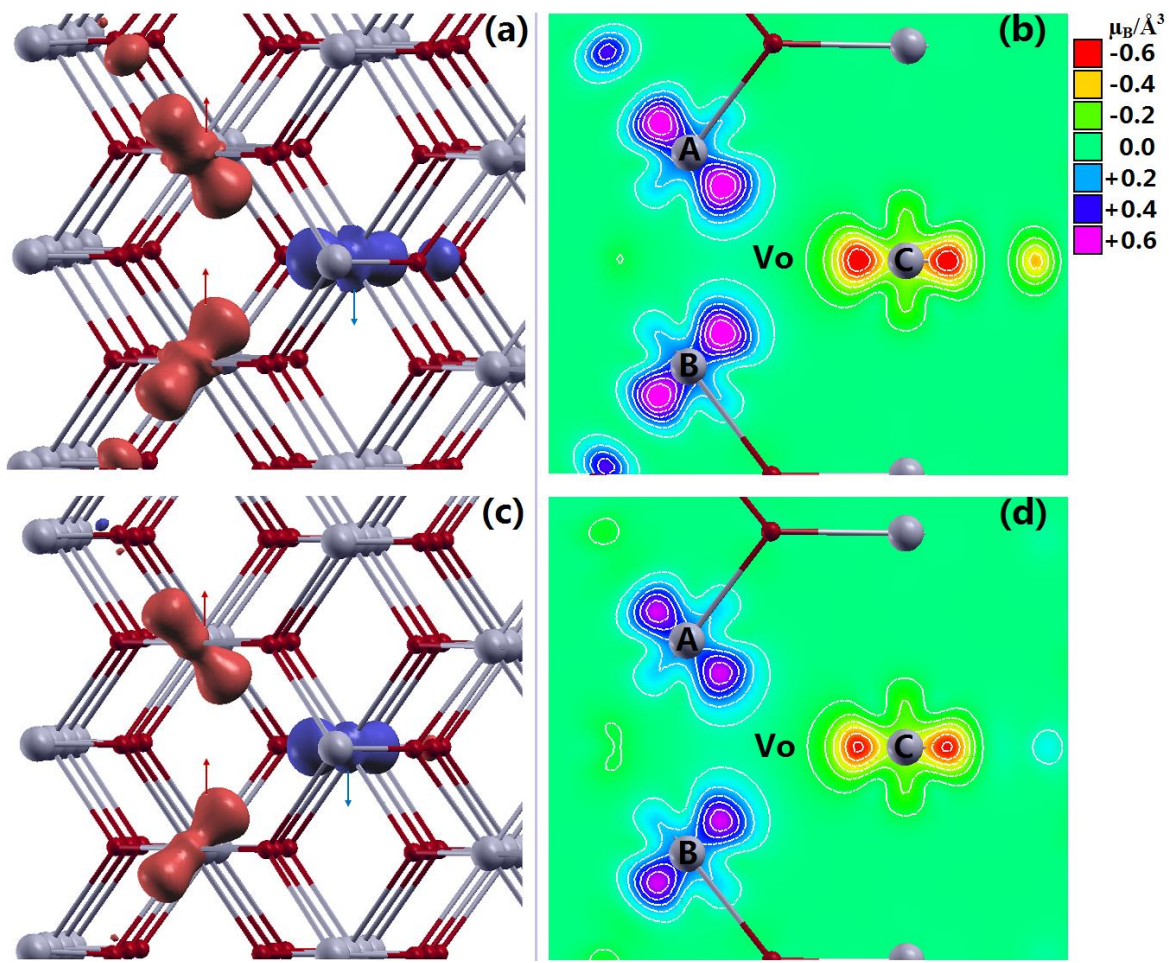

**FIG. 11** (color online) Similar to Fig. 6, but for the PtO$_{1.986}$ system at charge states $\delta q = -1e$ (upper panels (a), (b)) and $\delta q = +1e$ (lower panels (c), (d)).